\documentclass[12pt]{iopart}
\usepackage{graphicx,bm}

\begin{document}

\title[Conformal Cosmology]{Conformal Cosmology with a Positive \\
Effective Gravitational Constant }

\author[Peter R. Phillips]{Peter R. Phillips \\
Department of Physics, Washington University, St. Louis, MO 63130 }
\date{\today}

\begin{abstract}
The conformal cosmological model presented by Mannheim predicts a negative
value for the effective gravitational constant, $G_{\rm eff}$. It also
involves a scalar field, $S$, which is treated classically. In this paper
we point out that a classical treatment of $S$ is inappropriate, because
the Hamiltonian is non-Hermitean, and the theory must be developed in the
way pioneered by Bender and others. When this is done, we arrive at a
Hamiltonian with an energy spectrum that is bounded below, and also a
$G_{\rm eff}$ that is positive. The resulting theory closely resembles
the conventional cosmology based on Einstein relativity.
\end{abstract}



\maketitle


\section{INTRODUCTION}
\label{sec:intro}

Mannheim \cite{mann1} (henceforth M6) has developed a cosmological model
(MM) based on conformal invariance. This model has a number of attractive
features, but has not so far been accepted as a viable candidate for the
correct theory of gravitation, in large part because it predicts,
apparently unambiguously, a negative value for the effective gravitational
constant, $G_{\rm eff}$.  From this follows a thermal history markedly
different from the usual one, and consequent difficulties in addressing
questions such as primordial nucleosynthesis \cite{elye}.

In this paper we point out a mathematical problem in the formulation of the
model, and present a new approach that yields a positive $G_{\rm eff}$, so
that the model closely resembles the conventional one.

\section{Basic equations of the Mannheim model}

The MM involves a scalar field, $S$, in a Friedmann-Robertson-Walker (FRW)
background metric. It is this field $S$ that we will be concentrating on in
this paper. The treatment of both $S$ and the gravitational field in M6 is
classical. Important subsequent work has been done on the quantization of
the fourth-order equations for the gravitational field that result from the
conformal Lagrangian \cite{mann4}. The scalar field, $S$, however, has
always been treated classically, a procedure that we will question in this
paper.

There are several things that are required of $S$ in a viable model:
\newcounter{change}
\begin{list}{$\bullet$ \arabic{change}}{\usecounter{change}%
\setlength{\leftmargin}{0.25in}%
\setlength{\rightmargin}{0.25in}
}
\item An equation of motion that is conformally invariant.
\item An energy-momentum tensor whose 00 component (Hamiltonian) has
an energy spectrum that is bounded below.
\item A constant vacuum expectation value, $S_0$, derived from the equation
of motion. The $S_{0}^{2}$ in the action generates the effective
gravitational constant, $G_{\rm eff}$.
\item A $S_{0}^{4}$ term in the action that gives the correct sign for the
cosmological constant.
\end{list}

Let us see whether these criteria are met in the MM.
Using (as Mannheim does) a metric signature $-,\;+,\;+,\;+$, and
neglecting the coupling to the fermion field, the terms in Mannheim's
Lagrangian involving $S$ are (M6 (61)):
\begin{equation}
{\cal L} = - (-g)^{1/2} \left(\frac{1}{2} S^{;\mu} S_{;\mu}
- \frac{1}{12} S^2 R^{\mu}_{\;\;\mu} + \lambda_{\rm M} S^4 \right)
\label{eq:L}
\end{equation}

This results in an equation of motion (M6 (63)):
\begin{equation}
S^{;\mu}_{\;\; ;\mu} + \frac{1}{6} S R^{\mu}_{\;\;\mu}
- 4 \lambda_{\rm M} S^3 = 0
\label{eq:S}
\end{equation}
This equation, as required, is conformally invariant.

$S$ is assumed to acquire a non-zero vacuum expectation value, $S_0$,
found by setting the derivative term in (\ref{eq:S}) equal to zero.
This gives
\begin{equation}
S_{0}^{2} = \frac{R^{\mu}_{\;\;\mu}}{24 \lambda_{\rm M}}
\label{eq:Svev}
\end{equation}
as in \cite{mann2} (13), with $h=0$.

Further development of the model (M6, section 10) shows that,
in order to meet observational criteria at the present time,
$R^{\mu}_{\;\;\mu}$ and $\lambda_{\rm M}$ are both negative, so that
$S_{0}^{2}$ is real and positive. The effective gravitational constant,
$G_{\rm eff}$, in the MM turns out to be negative (M6 (224)):
\begin{equation}
G_{\rm eff} = - \frac{1}{4 \pi S_{0}^{2}}
\label{eq:Geff}
\end{equation}
choosing units with $c=1$.

The stress tensor in the MM model is obtained by taking the variation of
the action with respect to the metric, in the usual way (M6 (64); a similar
expression is given in \cite{birrell}). Retaining just the terms involving
$S$ we get
\begin{eqnarray}
T_{\mu \nu} & = & \frac{2}{3} S_{;\mu} S_{;\nu}
- \frac{1}{6} g_{\mu \nu} S S^{;\alpha}_{\;\; ;\alpha}
- \frac{1}{3} S S_{;\mu;\nu}
+ \frac{1}{3} g_{\mu \nu} S S^{;\alpha}_{\;\; ;\alpha}  
\nonumber \\
  & & {} - \frac{1}{6} S^2 \left( R_{\mu \nu}
- \frac{1}{2} g_{\mu \nu} R^{\alpha}_{\;\;\alpha} \right)
- g_{\mu \nu} \lambda_{M} S^4
\label{eq:T_Mann}
\end{eqnarray}

The Lagrangian (\ref{eq:L}) is analogous to the Minkowski space
Lagrangian for a $\phi^4$ field theory (\cite{wein2} (7.2.14), with
${\cal H} (\phi) = \mu^2 \phi^4$):
\begin{equation}
{\cal L}_{\rm mink} = \frac{1}{2} (\partial_t \phi)^2
- \frac{1}{2} (\nabla \phi)^2 - \frac{m^2}{2} \phi^2
- \mu^2 \phi^4
\label{eq:M}
\end{equation}

The $S^2$ term in (\ref{eq:L}) has the ``right'' sign for an $m^2$ term. But
with $\lambda_M < 0$, the $S^4$ term has the ``wrong'' sign, and, in a
conventional treatment, will lead to a spectrum that is not bounded below.
A theory of this sort has to be treated by methods appropriate to
non-Hermitean Hamiltonians \cite{bender2}. At the level of
quantum mechanics, ``wrong sign'' $\phi^4$ theory is well developed.
Nevertheless, even in Minkowski space, constructing a corresponding quantum
field theory is a difficult problem, still incompletely understood
\cite{bender3}. Conformal cosmology will remain flawed until we can make
progress in understanding the scalar field.

\section{A new approach to non-Hermitean $\phi^4$ theory in Minkowski space}

A conformally invariant theory with a single scalar field has a unique
action, M6 (61), provided we use the familiar techniques
appropriate for Hermitean Lagrangians. Once we recognize that our theory
involves a non-Hermitean Lagrangian, however, a new approach is suggested,
that we introduce in this section. We begin by working in Minkowski space,
but retain $g_{\mu \nu}$ and $(-g)^{1/2}$ in formulae to simplify the
transition to a FRW space.

A $\phi^4$ theory with a ``wrong sign'' $\phi^4$ term is non-Hermitean but
is nevertheless ${\cal PT}$ symmetric, and can be treated by the methods
outlined in \cite{bender2}. The distinctive feature of this approach is the
use of the ${\cal CPT}$ norm in place of the usual Dirac norm; for a
quantized field, ${\mathsf S}$, we write this norm as
\begin{equation}
N({\mathsf S}) = \langle | {\mathsf S}^{\cal CPT} {\mathsf S} | \rangle
\label{eq:normex}
\end{equation}
Here ${\cal P}$ and ${\cal T}$ represent the usual parity and time-reversal
operations, while ${\cal C}$ represents a special operation designed to
ensure the norm is real and positive definite and the theory is unitary. The
${\cal C}$ operator has to be specifically calculated for each Hamiltonian.

Our cosmological model is written in terms of classical fields (expectation
values, $S(x)$), which we take to be real. We assume $S^{\cal CPT} (x)$
can then be expressed as
\begin{eqnarray}
S^{\cal CPT} (x) & \equiv &
\int {\rmd}^4 y \, C(x^{\mu} - y^{\mu}) S^{*} (-y) \nonumber \\
  & = & \int {\rmd}^4 y \, C(x^{\mu} - y^{\mu})
\left[S^{*} (z) \right]_{\forall \rho : z^{\rho} = -y^{\rho}} 
\label{eq:CPTdef}
\end{eqnarray}
(compare \cite{bender2} (78)), so that
\begin{eqnarray}
N(S) & = & \int {\rmd}^4 x \int {\rmd}^4 y \, C (x^{\mu} - y^{\mu} ) 
\left[ S^{*} (z) \right]_{\forall \rho : z^{\rho} = -y^{\rho}}
S (x) \label{eq:normex2}
\end{eqnarray}
In an expression of this sort, $S (x)$ and $S^{*} (-y)$ describe the field
$S$ at the same physical point, but use different coordinate systems to
refer to that point.

Take the complex conjugate of (\ref{eq:normex2}), and let
$x^{\mu} \rightarrow -y^{\mu}$ and $y^{\mu} \rightarrow -x^{\mu}$:
\begin{eqnarray}
N(S) & = & \int {\rmd}^4 x \int {\rmd}^4 y \, C^{*} (x^{\mu} - y^{\mu} ) 
\left[ S(z) \right]_{\forall \rho : z^{\rho} = x^{\rho}}
\left[ S^{*} (u) \right]_{\forall \rho : u^{\rho} = -y^{\rho}} \nonumber \\
  & = & \int {\rmd}^4 x \int {\rmd}^4 y \, C^{*} (x^{\mu} - y^{\mu} )
S^{*} (-y) S(x)
\end{eqnarray}
showing that $C(x^{\mu} - y^{\mu})$ must be real.

\subsection{The action}

We define our action by
\begin{eqnarray}
I & \equiv & \int {\rmd}^4 x \, (-g)^{1/2} \left\{
\frac{\sigma_{k} }{2} g^{\mu \nu}
\left[ \frac{\partial S (x)}{\partial x^{\mu}} \right]^{\cal CPT}
\frac{\partial S (x)}{\partial x^{\nu}} \right. \nonumber \\
  & & \left. {} + \frac{\sigma_m}{2} m^2 S^{\cal CPT}(x) S(x)
+ \sigma_{\mu} {\mu}^2 \left[ S^{\cal CPT}(x) S(x) \right]^2
\right\} \label{eq:Imink} 
\end{eqnarray}
where $\sigma_k$, $\sigma_m$ and $\sigma_{\mu}$ are simply ``sign factors'',
each of which can be equal to $\pm 1$. We will determine their actual values
as we proceed.

\subsection{Energy-momentum tensor}

The energy-momentum tensor is obtained in the usual way by varying the
action with respect to the metric:
\begin{eqnarray}
T^{\mu \nu} (x) & \equiv & \frac{2}{\left( -g \right)^{1/2} } 
\frac{\delta I}{\delta g_{\mu \nu} (x)} 
\label{eq:Tdef} \nonumber \\
  & = & \frac{2}{\left( -g \right)^{1/2} } 
\left( \frac{1}{2 ( -g )^{1/2}} (-g) \,
g^{\mu \nu} \left\{ \frac{\sigma_k}{2} g^{\alpha \beta}
\left[ \frac{\partial S (x)}{\partial x^{\alpha}} \right]^{\cal CPT}
\frac{\partial S (x)}{\partial x^{\beta}} \right. \right. \nonumber \\
  & & \left. {} + \frac{\sigma_m}{2} m^2 S^{\cal CPT} (x) S(x) 
+ \sigma_{\mu} {\mu}^2 \left[ S^{\cal CPT} (x) S(x) \right]^2 \right\}
\nonumber \\
  & & \left. {} - (-g )^{1/2} \frac{\sigma_k}{2} g^{\mu \alpha} g^{\nu \beta}
\left[ \frac{\partial S (x)}{\partial x^{\alpha}} \right]^{\cal CPT}
\frac{\partial S (x)}{\partial x^{\beta}} \right) \nonumber  \\
& = & g^{\mu \nu}
\left\{ \frac{\sigma_k}{2} g^{\alpha \beta}
\left[ \frac{\partial S (x)}{\partial x^{\alpha}} \right]^{\cal CPT}
\frac{\partial S (x)}{\partial x^{\beta}} \right.  \nonumber \\
  & & \left. {} + \frac{\sigma_m}{2} m^2 S^{\cal CPT} (x) S(x)
+ \sigma_{\mu} {\mu}^2 \left[ S^{\cal CPT} (x) S(x) \right]^2 \right\}
\nonumber \\
  & & {} - \sigma_k g^{\mu \alpha} g^{\nu \beta}
\left[ \frac{\partial S (x)}{\partial x^{\alpha}} \right]^{\cal CPT}
\frac{\partial S (x)}{\partial x^{\beta}}
\end{eqnarray}

Setting $g_{\mu \nu} = {\rm diag\;} (-1, +1, +1, +1 )$, the Hamiltonian is
given by
\begin{eqnarray}
{\cal H} & \equiv & T^{00} \nonumber \\
  & = & - \left( \frac{\sigma_k}{2} g^{\alpha \beta}
\left[ \frac{\partial S (x)}{\partial x^{\alpha}}
\right]^{\cal CPT}
\frac{\partial S(x)}{\partial x^{\beta}} \right.
+ \frac{\sigma_m}{2} m^2 \left[ S(x) \right]^{\cal CPT} S(x)
\nonumber \\
  & & \left. {} + \sigma_{\mu} {\mu}^2 \left\{
\left[ S(x) \right]^{\cal CPT} S(x) \right\}^2 \right)
- \sigma_k \left[ \frac{\partial S(x)}{\partial x^{0}} \right]^{\cal CPT}
\frac{\partial S(x)}{\partial x^{0}}  \nonumber \\
  & = & - \frac{\sigma_k }{2}
\left[ \frac{\partial S(x)}{\partial x^{0}} \right]^{\cal CPT}
\frac{\partial S(x)}{\partial x^{0}} - \frac{\sigma_k }{2}
\left[ \nabla S(x) \right]^{\cal CPT} \nabla S(x) \nonumber \\
  & & {} - \frac{\sigma_m}{2} m^2 \left[ S(x) \right]^{\cal CPT} S(x)
- \sigma_{\mu} {\mu}^2 \left\{
\left[ S(x) \right]^{\cal CPT} S(x) \right\}^2  
\label{eq:ham}
\end{eqnarray}

All four terms in (\ref{eq:ham}) include a (positive definite) ${\cal CPT}$
norm. By analogy with ordinary $\phi^4$ theory, the first two terms and the
last must be positive if we are to have an energy spectrum that is bounded
below. We therefore set $\sigma_k = \sigma_{\mu} = -1$.

The sign of the $\sigma_m$ term will be determined by the requirement of
conformal invariance when we go to FRW space.

\subsection{Equation of motion}

The equation of motion for $S$ is obtained in the usual way, by requiring
the action to be stationary under small variations, $\delta S$. The
appearance of $S^{\cal CPT}$ in the action is unusual, and the variation
requires some care; details are given in the appendix. The
equation of motion is given in (\ref{eqapp:motion}). Writing
$\sigma_k = \sigma_{\mu} = -1$, this becomes
\begin{equation}
g^{\mu \nu} \frac{\partial^2 S(x)}{\partial x^{\mu} \partial x^{\nu}} 
- \sigma_m m^2 S(x) 
+ 4 \mu^2 \left[ S(x) S^{\cal CPT} (x) \right] S(x) = 0
\label{eq:motion}
\end{equation}

Comparing this with Mannheim's equation of motion, (\ref{eq:S}), we
infer that we go over to FRW space by setting
$\sigma_m m^2 = -R^{\mu}_{\;\;\mu}/6$. Since $R^{\mu}_{\;\;\mu} < 0$, we
must set $\sigma_m = +1$. To maintain Mannheim's notation as far as
possible, we will here define $\lambda_{\rm M} = -\mu^2 < 0$, giving the
equation of motion
\begin{equation}
g^{\mu \nu} \frac{\partial^2 S(x)}{\partial x^{\mu} \partial x^{\nu}} 
+ \frac{1}{6} R^{\mu}_{\;\;\mu} S(x) 
- 4 \lambda_{\rm M} \left[ S(x) S^{\cal CPT} (x) \right] S(x) = 0
\label{eq:motion2}
\end{equation}

Following Mannheim, we assume that $S$ develops a constant vacuum
expectation value, calculated by setting the derivative term in
(\ref{eq:motion2}) equal to zero. The result is the same as (\ref{eq:Svev}),
$S_{0}^{2} = R^{\mu}_{\;\;\mu} / 24 \lambda_{\rm M}$, where,
as before, $R^{\mu}_{\;\;\mu} < 0$ and $\lambda_{\rm M} < 0$.

\subsection{The action and energy-momentum tensor revisited}

We can now use (\ref{eq:Imink}) to write the action in FRW space:
\begin{eqnarray}
I & \equiv & \int {\rmd}^4 x \, (-g)^{1/2} \left\{
-\frac{1}{2} g^{\mu \nu}
\left[ S (x) \right]^{\cal CPT}_{;\mu}
\left[S (x) \right]_{;\nu} \right. \nonumber \\
  & & \left. {} - \frac{R^{\sigma}_{\;\;\sigma}}{12} S^{\cal CPT}(x) S(x)
+ \lambda_{\rm M} \left[ S^{\cal CPT}(x) S(x) \right]^2
\right\} \label{eq:IFRW} 
\end{eqnarray}

From this we can infer the energy-momentum tensor analogous to
(\ref{eq:T_Mann}). For a cosmological model comparable to Mannheim's
only the last two terms are of interest, which are
\begin{equation}
T^{\mu \nu} \approx \frac{1}{6} \left( R^{\mu \nu}
- \frac{1}{2} g^{\mu \nu} R^{\alpha}_{\;\;\alpha} \right)
\left[ S^{\cal CPT}(x) S(x) \right]
+ g^{\mu \nu} \lambda_{M}
\left[ S^{\cal CPT}(x) S(x) \right]^2
\label{eq:T_Mann2}
\end{equation}

Replacing $S(x)$ by its vacuum expectation value, $S_0$, we get
\begin{equation}
T^{\mu \nu} \approx \frac{1}{6} \left( R^{\mu \nu}
- \frac{1}{2} g^{\mu \nu} R^{\alpha}_{\;\;\alpha} \right) S_{0}^{2}
+ g^{\mu \nu} \lambda_{M} S_{0}^{4}
\label{eq:T_Mann3}
\end{equation}

The important point is that the signs of both these terms are reversed in
comparison to (\ref{eq:T_Mann}).

\section{Cosmological implications}

Mannheim points out (M6, section 10) that because the FRW space is
conformally flat, the cosmological equation of motion (CEM) reduces to
\begin{equation}
T^{\mu \nu}_{\rm total} = 0
\label{eq:motion3}
\end{equation}

$T^{\mu \nu}_{\rm total}$ is just the sum of (\ref{eq:T_Mann3}) and
$T^{\mu \nu}_{\rm kin}$, the contribution from ordinary matter (fermion
fields, electromagnetic radiation, etc.). So our CEM, analogous to
M6 (222), becomes
\begin{equation}
T^{\mu \nu}_{\rm total} = T^{\mu \nu}_{\rm kin} 
+ \frac{1}{6} S_{0}^{2} \left( R^{\mu \nu}
- \frac{1}{2} g^{\mu \nu} R^{\alpha}_{\;\;\alpha} \right)
+ g^{\mu \nu} \lambda_{\rm M} S_{0}^{4} = 0
\label{eq:motion4}
\end{equation}
or, as in M6 (223)
\begin{equation}
\frac{1}{6} S_{0}^{2} \left( R^{\mu \nu}
- \frac{1}{2} g^{\mu \nu} R^{\alpha}_{\;\;\alpha} \right)
  = - T^{\mu \nu}_{\rm kin} - g^{\mu \nu} \Lambda_{\rm M}
\label{eq:motion5}
\end{equation}
with $\Lambda_{\rm M} \equiv \lambda_{\rm M} S_{0}^{4} < 0$.

The $S_{0}^{2}$ term defines the effective gravitational constant in the
theory. We find
\begin{equation}
G_{\rm eff} = \frac{3}{4 \pi S_{0}^{2}}
\label{eq:Gdef2}
\end{equation}
analogous to M6 (224), but wth a positive sign.

Of particular interest, since it permits an analytic solution, is a model
containing radiation only (M6, section 10.2). A corresponding solution
exists for the present model also; let us see how it differs. The equation
analogous to M6 (226) is (with $H \equiv \dot{R}/R$):
\begin{equation}
\dot{R}^{2} (t)  =  \dot{R}^2 (t) \left( \overline{\Omega}_{M} (t)
+ \overline{\Omega}_{\Lambda} (t) + \overline{\Omega}_{k} (t) \right) 
\end{equation}
\begin{eqnarray}
\overline{\Omega}_M (t) & = & \frac{8 \pi G_{\rm eff} \rho}{3 H^2} \\
\overline{\Omega}_{\Lambda} (t) & = &
- \frac{8 \pi G_{\rm eff} \Lambda_M}{3 H^2} \\
\overline{\Omega}_{k} (t) & = &
- \frac{k}{\dot{R}^2 (t)} 
\end{eqnarray}
\begin{equation}
\overline{\Omega}_M (t) + \overline{\Omega}_{\Lambda} (t) 
+ \overline{\Omega}_{k} (t) = 1
\label{eq:Omega}
\end{equation}
Since $k < 0$ for the open geometry of the MM, all three terms in
(\ref{eq:Omega}) are positive, in contrast to M6, where
$\overline{\Omega}_M$ is negative.

Write the solution as in M6 (230), with $\rho_{M} (t) = A/R^{4}$:
\begin{equation}
R^{2} (t) = - \frac{ k (\beta - 1)}{2 \alpha}
- \frac{ k \beta \sinh^2 ( \alpha^{1/2} t)}{\alpha}
\label{eq:R}
\end{equation}
\begin{eqnarray}
\alpha & = & - \frac{ 8 \pi G_{\rm eff} \Lambda_M }{3} > 0
\label{eq:alpha} \\
\beta & = & \left( 1 + \frac{16 A \lambda_{\rm M} }{k^2} \right)^{1/2} < 1
\label{eq:beta} 
\end{eqnarray}

\begin{figure}
\centering
\includegraphics[scale=0.85]{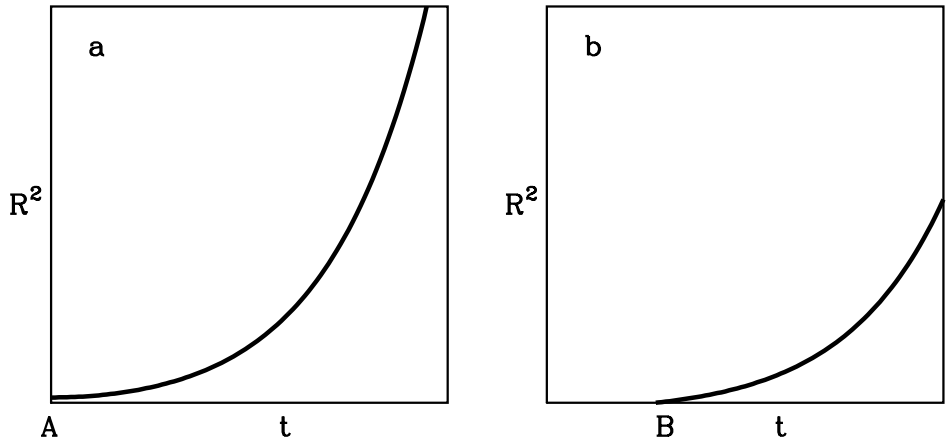}
\caption{\label{fig:ccpgfigure1a_1b}
Two curves drawn using (\ref{eq:R}), but with
different values of parameters. (a) $k=-1$, $\alpha=1.0$, $\beta=1.4$. 
$R$ starts from a non-zero value at $t=0$ (point A). (b) $k=-1$,
$\alpha=1.0$, $\beta=0.6$. Part of the curve now lies below the horizontal
axis, and is non-physical. The origin of $t$ is conventionally shifted to
B, where $R=0$. }
\end{figure}

The first term on the right of (\ref{eq:R}) is negative, not positive as in
the MM. This means that $R$ must start from zero, as in the standard
cosmology, not from some finite value, as in the MM. This is illustrated in
figure \ref{fig:ccpgfigure1a_1b}.  Both curves were drawn using the formula
(\ref{eq:R}), but in the upper graph $\beta > 1$, while in the lower graph
$\beta < 1$. The origin of $t$ is conventionally shifted in the lower graph
so that $t=0$ is at point B, where $R=0$.

\section{How does the new model differ from conventional cosmology?}

The present model has two features that are not present in conventional
cosmology based on Einstein's equations.

First, the model includes a scalar field that is essentially massless.
The non-zero vacuum expectation value of this field is essential, but we
have ignored any excitations. This may be permissible because the field
couples very weakly to normal matter, and is difficult to observe, or it
may undergo a spontaneous transition that renders it massive.

Second, Mannheim uses (\ref{eq:motion3}) for his basic cosmological equation,
rather than the more complete one that results from the conformal action,
M6 (188):
\begin{equation}
T^{\mu \nu}_{\rm total} = 4 \alpha_g W^{\mu \nu}
\label{eq:motion6}
\end{equation}
where $W^{\mu \nu}$ is the Weyl tensor, defined in M6 (107), (108) and
(185). Mannheim justifies the neglect of $W^{\mu \nu}$ by observing that
a FRW metric is conformally flat, and in such a space $W^{\mu \nu} = 0$.
This is true, but if we are to include perturbations to the metric (as,
for example, in the study of anisotropies of the CMB) then this neglected
term may become important.

\section{Conclusion}
\label{sec:conclude}

We have pointed out two flaws in Mannheim's conformal cosmological model.

\setcounter{change}{0}
\begin{list}{$\bullet$ \arabic{change}}{\usecounter{change}%
\setlength{\leftmargin}{0.25in}%
\setlength{\rightmargin}{0.25in}
}
\item The model predicts, apparently unambiguously, a negative value for the
effective gravitational constant, $G_{\rm eff}$.
\item The model involves a scalar field, $S (x)$, that satisfies a
conformally invariant equation of motion and develops a vacuum expectation
value, $S_0$. The values of the parameters that are needed to satisfy
observations lead to a ``wrong sign $S^4$ theory'', with a Hamiltonian that
has a spectrum that is unbounded below.
\end{list}

We have attempted to apply the techniques appropriate for such Hamiltonians
to this cosmological problem, restricting ourselves to the classical
limit of field equations that are still imperfectly understood. In this
limit, using assumptions that appear reasonable, we find both a positive
value for $G_{\rm eff}$ and a positive definite spectrum for the Hamiltonian.

Our derivation depends on one simple observation, the change of sign as we
go from (\ref{eq:deltaIka}) to (\ref{eq:deltaIkb}).

The derivation presented here will remain conjectural until progress is
made in two main directions:
\setcounter{change}{0}
\begin{list}{$\bullet$ \arabic{change}}{\usecounter{change}%
\setlength{\leftmargin}{0.25in}%
\setlength{\rightmargin}{0.25in}
}
\item The techniques that have been successfully applied to $\phi^4$ quantum
mechanics will have to be developed to cover the corresponding quantum field
theory; this seems not to have been achieved at this time \cite{bender3}.
In particular, the form (\ref{eq:CPTdef})  for $S^{\cal CPT} (x)$ must be
shown to be appropriate at the classical level.
\item The various manipulations we have employed are suitable for Minkowski
space, but more detailed investigations are needed to show whether they can
legitimately be extended to a FRW space.
\end{list}

If, on the other hand, we can accept the present model as viable, without
first filling in these important gaps in our understanding, then we have
to face the difficult question: how can we conclusively distinguish this
model from the conventional one?

\ack

The department of physics at Washington University, in particular Carl
Bender, have given invaluable support to this retired colleague.

\appendix
\section*{Appendix: variation of the action}
\label{appsec:action}
\setcounter{section}{1}

The action is defined in (\ref{eq:Imink}). We will start with the simplest
term,
\begin{eqnarray}
\fl I_m & \equiv & \int {\rmd}^4 x \, (-g)^{1/2}
\frac{\sigma_m}{2} m^2 S^{\cal CPT}(x) S(x) \nonumber \\
\fl  & = & \int {\rmd}^4 x \int {\rmd}^4 y \, (-g)^{1/2}
\frac{\sigma_m}{2} m^2 C(x^{\mu} - y^{\mu} )
\left[ S^{*}(z) \right]_{\forall \rho : z^{\rho} = -y^{\rho}}
S(x) \label{eq:Ima}
\end{eqnarray}

Let $S(x)$ vary by a small amount $\delta S (x)$. Then
$S^{*} (-x)$ will vary by $\delta \left[S^{*} (-x)\right]$.
The variation of $I_m$ will be the sum of two terms, $\delta I_m (1)$ and
$\delta I_m (2)$:
\begin{equation}
\fl \delta I_m (1) = \int {\rmd}^4 x \int {\rmd}^4 y \,
(-g)^{1/2} \frac{\sigma_m}{2} m^2 C(x^{\mu} - y^{\mu} )
\left[ S^{*}(z) \right]_{\forall \rho : z^{\rho} = -y^{\rho}}
\delta S(x) \label{eq:deltaIm1} 
\end{equation}
\begin{equation}
\fl \delta I_m (2) = \int {\rmd}^4 x \int {\rmd}^4 y \,
(-g)^{1/2} \frac{\sigma_m}{2} m^2 C(x^{\mu} - y^{\mu} )
\delta \left[ S^{*}(z)
\right]_{\forall \rho : z^{\rho} = -y^{\rho}} S(x) \label{eq:deltaIm2}
\end{equation}

Take the complex conjugate of (\ref{eq:deltaIm2}), and let 
$x^{\mu} \rightarrow -y^{\mu}$, $y^{\mu} \rightarrow -x^{\mu}$:
\begin{eqnarray}
\delta I^{*}_{m} (2) & = & \int {\rmd}^4 x \int {\rmd}^4 y \, (-g)^{1/2}
\frac{\sigma_m}{2} m^2 C(x^{\mu} - y^{\mu} )
\delta S (x) S^{*} (-y) \nonumber \\
  & = & \delta I_m (1)
\end{eqnarray}
so that
\begin{eqnarray}
\fl \delta I_m & = & 2 {\rm Re}\left[ \delta I_m (1) \right]
\nonumber \\
\fl & = & \int {\rmd}^4 x \int {\rmd}^4 y \, (-g)^{1/2}
\sigma_m m^2 C(x^{\mu} - y^{\mu} )
{\rm Re} \left\{\left[ S^{*}(z)
\right]_{\forall \rho : z^{\rho} = -y^{\rho}}
\delta S(x) \right\} \label{eq:deltaImb}
\end{eqnarray}
 
The $\mu^2$ term in the action can be treated in the same way, to give
\begin{eqnarray}
\fl \delta I_{\mu} & = & 4 \int {\rmd}^4 x \int {\rmd}^4 y \, ( -g )^{1/2}
\sigma_{\mu} \mu^2 C( x^{\mu} - y^{\mu} ) 
\nonumber \\
\fl  & & \times \, \left[ S^{\cal CPT} (x) S(x) \right] {\rm Re} \left\{
\left[ S^{*} (z) \right]_{\forall \rho : z^{\rho} = - y^{\rho}}
\delta S (x) \right\}
\end{eqnarray}

Calculating the variation of the kinetic term starts just as with
conventional Lagrangians. We imagine a variation
$\delta \left( \partial S (x) / \partial x^{\nu} \right)$, and convert
this to a variation of $S (x)$ by an integration by parts, discarding a
surface term.  Recalling that we are working in Minkowski space, where the
metric tensor is constant, we get
\begin{equation}
\fl \delta I_k (1) = - \int {\rmd}^4 x \, (-g)^{1/2}
\frac{\sigma_k }{2} g^{\mu \nu}
\frac{ \partial }{\partial x^{\nu}}
\left[ \frac{\partial S (x) }{ \partial x^{\mu}} \right]^{\cal CPT}
\delta S (x)
\end{equation}

Using (\ref{eq:CPTdef}) we write this as
\begin{eqnarray}
\fl \delta I_k (1) & = & - \int {\rmd}^4 x \int {\rmd}^4 y \,
(-g)^{1/2} \frac{\sigma_k }{2} g^{\mu \nu}
\frac{ \partial }{\partial x^{\nu}}
C(x^{\mu} - y^{\mu}) \nonumber \\
\fl  & & \times \, \left[
\frac{\partial S^{*} (z)}{\partial z^{\mu}}
\right]_{\forall \rho : z^{\rho} = -y^{\rho}} \delta S (x) \\
\fl & = & \int {\rmd}^4 x \int {\rmd}^4 y \, (-g)^{1/2}
\frac{\sigma_k }{2} g^{\mu \nu}
\left[ \frac{ \partial }{\partial y^{\nu}}
C(x^{\mu} - y^{\mu}) \right] \nonumber \\
\fl  & & \times \, \left[
\frac{\partial S^{*} (z)}{\partial z^{\mu}}
\right]_{\forall \rho : z^{\rho} = -y^{\rho}} \delta S (x)
\end{eqnarray}

Now integrate by parts and discard a surface term:
\begin{eqnarray}
\fl \delta I_k (1) & = & - \int {\rmd}^4 x \int {\rmd}^4 y \,
(-g)^{1/2} \frac{\sigma_k }{2} g^{\mu \nu} 
C(x^{\mu} - y^{\mu}) \nonumber \\
\fl  & & \times \, \frac{ \partial }{\partial y^{\nu}}
\left[ \frac{\partial S^{*} (z)}{\partial z^{\mu}}
\right]_{\forall \rho : z^{\rho} = -y^{\rho}}
\delta S (x) \label{eq:deltaIka} \\
\fl & = & \int {\rmd}^4 x \int {\rmd}^4 y \, (-g)^{1/2}
\frac{\sigma_k }{2} g^{\mu \nu}
C(x^{\mu} - y^{\mu})
\left[ \frac{\partial^2 S^{*} (z)}{\partial z^{\mu} \partial z^{\nu}}
\right]_{\forall \rho : z^{\rho} = -y^{\rho}}
\delta S (x) \label{eq:deltaIkb} 
\end{eqnarray}

The passage from (\ref{eq:deltaIka}) to (\ref{eq:deltaIkb}) is best
illustrated by an example. Suppose
\begin{equation}
\frac{\partial S^{*} (x) }{ \partial x^{\mu}} =
\left( k_{\sigma} x^{\sigma} \right)^n k_{\mu}
\end{equation}
 where $k_{\mu}$ is some fixed vector. Then
\begin{equation}
\frac{\partial}{\partial x^{\nu}}
\left[ \frac{ \partial S^{*} (y) }{\partial y^{\mu}}
\right]_{\forall \rho : y^{\rho} = - x^{\rho}} = 
 n (-1)^{n} \left( k_{\sigma } x^{\sigma} \right)^{n-1} k_{\mu} k_{\nu}
\label{eqapp:difex}
\end{equation}
and
\begin{eqnarray}
\left[\frac{\partial^2 S^{*} (y)}{\partial y^{\nu} \partial y^{\mu}}
\right]_{\forall \rho : y^{\rho} = -x^{\rho}} & = &
\left[ n \left( k_{\sigma} y^{\sigma} \right)^{n-1} k_{\mu} k_{\nu}
\right]_{\forall \rho : y^{\rho} = - x^{\rho}} \nonumber \\
  & = &
n (-1)^{n-1} \left( k_{\sigma} x^{\sigma} \right)^{n-1} k_{\mu} k_{\nu}
\label{eqapp:difex2}
\end{eqnarray}

Note the change in sign from (\ref{eqapp:difex}) to (\ref{eqapp:difex2});
it implies that the variation of the kinetic term results in
\begin{equation}
\fl \delta I_k = {\rm Re}
\left\{ \int {\rmd}^4 x \int {\rmd}^4 y \, (-g)^{1/2}
\sigma_k g^{\mu \nu} C(x^{\mu} - y^{\mu})
\left[ \frac{\partial^2 S^{*} (z)}{\partial z^{\mu} \partial z^{\nu}}
\right]_{\forall \rho : z^{\rho} = -y^{\rho}} \delta S (x)
\right\} \label{eq:deltaIkc} 
\end{equation}

Now use $\delta I_m + \delta I_{\mu} + \delta I_k = 0$ for arbitrary
(possibly complex) variations $\delta S(x)$ to get the equation of motion
\begin{equation}
\fl \sigma_k g^{\mu \nu}
\frac{\partial^2 S(x)}{\partial x^{\mu} \partial x^{\nu}} 
+ \sigma_m m^2 S(x)
+ 4 \sigma_{\mu} \mu^2 \left[ S^{\cal CPT} (x) S(x) \right] S(x) = 0
\label{eqapp:motion}
\end{equation}

\end{document}